\documentclass[authoryear,round,12pt]{article}
\usepackage{color}
\usepackage{pslatex}
\usepackage{graphicx}
\usepackage{setspace}
\usepackage{natbib}
\usepackage{amsmath}
\usepackage{subfigure}
\newcommand{\beq}{\begin{equation}}
\newcommand{\eeq}{\end{equation}}
\newcommand{\myabstract}{This document is another installment in a
  series of near real-time weekly influenza forecasts made during the
  2012-2013 influenza season.  Here we present some of the results of
  forecasts initiated following assimilation of observations for Week
  52 (i.e. the forecast begins December 30, 2012) for municipalities
  in the United States.  The forecasts were made on January 4, 2013.
  Results from forecasts initiated the five previous weeks (Weeks
  47-51) are also presented.}

\newcommand{\myacknow}{Funding was provided by US NIH grant GM100467
  (JS, AK, ML), as well as NIEHS Center grant ES009089 (JS) and
  the RAPIDD program of the Science and Technology Directorate, US
  Department of Homeland Security (JS).  The content is solely the
  responsibility of the authors and does not necessarily represent the
  official views of the National Institute Of General Medical
  Sciences, National Institutes of Health, or Department of Homeland
  Security.}

\begin{document}
%
%
\title{\textbf{\large{Week 52 Influenza Forecast for the 2012-2013
      U.S. Season}}}
%
%
\author{\textsc{Jeffrey Shaman}
                                \thanks{\textit{Corresponding author address:} 
                                Jeffrey Shaman, Department of
                                Environmental Health Sciences, Mailman
                                School of Public Health, Columbia
                                University, 722 West 168th Street,
                                Rosenfield Building, Room 1104C, New
                                York, NY 10032. 
                                \newline{E-mail:
                                  jls106@columbia.edu}}\quad\textsc{}\\
\centerline{\textit{\footnotesize{Department of Environmental Health Sciences,
    Mailman School of Public Health, Columbia University, New York, New York}}}
\and
\centerline{\textsc{Alicia Karspeck}} \\
\centerline{\textit{\footnotesize{Climate and Global Dynamics
      Division, National Center for Atmospheric Research, Boulder, Colorado}}}
\and 
\centerline{\textsc{Marc Lipsitch}} \\
\centerline{\textit{\footnotesize{Center for Communicable Disease
      Dynamics, Harvard School of Public Health, Harvard University,
      Boston, Massachussetts}}}
}

\maketitle

{
\begin{abstract}
\myabstract
\end{abstract}
}

\section{Observations}
\label{sec:obs}

The forecasts are again dealing with shifting observations.  Week 51
census division infectivity rates as posted this week (January 4,
2013) versus last week (when the Week 51 estimates were first
published on December 28, 2012) have changed substantially for a
number of divisions.  For instance, infectivity among assayed
individuals in the West South Central doubled to 26\% from 13\% with
the addition of much more data (1393 assays v. only 68 last week).
Clearly, the data last week were incomplete.  More typically the
division sample size increases 25-300\%; however, strangely, the
infectivity rate goes up in all but one instance (Table
\ref{table:t1}).  For New England, the infectivity rate increases 22
percentage points; for the Mid-Atlantic it doubles from 20\% to 40\%;
for the West North Central it increases 16 percentage points.

\begin{table}[t]
\caption{Week 51 Infectivity rates by Census Division as provided on
  December 28, 2012 and January 4, 2013.  Sample size is 
  given in parentheses.}\label{table:t1}
 \begin{center}
\resizebox{10cm}{!} {
 \begin{tabular}{ccccrrcrc}
   \hline\hline
   Region & December 28, 2012 & January 4, 2013\\
  \hline
 New England & 27.85\% (348) & 49.88\% (808)\\
 Mid-Atlantic & 20.04\% (484) & 40.11\% (1224)\\
 South Atlantic & 29.94\% (2498) & 34.01\% (3452)\\
 East North Central & 61.97\% (476) & 56.73\% (855)\\
 East South Central & 26.61\% (218) & 38.56\% (319)\\
 West North Central & 25.95\% (682) & 42.28\% (1147)\\
 West South Central & 13.24\% (68) & 26.49\% (1393) \\
 Mountain & 36.83\% (649) & 39.45\% (1985) \\
 Pacific & 15.54\% (811) & 16.68\% (1037)\\
   \hline
 \end{tabular}
}
 \end{center}
\end{table}

\begin{figure}[tbh]
\noindent\includegraphics[width=20pc,angle=0]{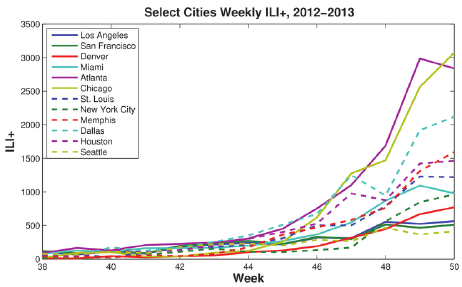}
\noindent\includegraphics[width=20pc,angle=0]{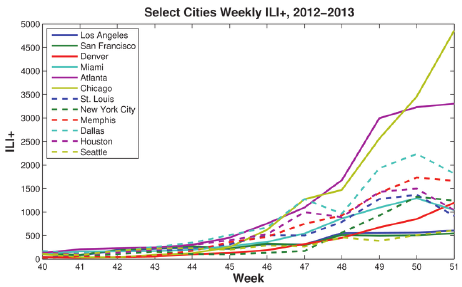}\\
\noindent\includegraphics[width=20pc,angle=0]{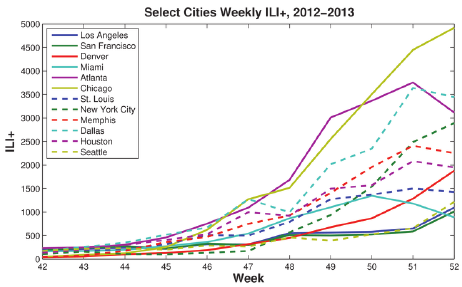}
\caption{Time series of: Top Left) Week 50 estimates of Weeks 38-50
  ILI+; Top Right) Week 51 estimates of Weeks 40-51 ILI+; and Bottom
  Left) Week 52 estimates of Weeks 42-52 ILI+ for the 2012-2013
  season.  ILI+ is Google Flu Trends weekly municipal ILI estimates
  times CDC census division seropositive rates.}
\label{fig:selcitytimeseries}
\end{figure}

These increases are large and affect the ILI+ observational metric.
Last week, based on the Week 51 observations, Dallas, Memphis and
Houston appeared to be abating, having peaked in Week 50 (Figure
\ref{fig:selcitytimeseries}).  This, of course, suggested to the model
during assimilation that the peak had abated as well, which likely
affected its parameter optimization and prediction trajectories.
However, this week, with the revised infectivity numbers, Week 51 is
now peak for these cities.

It is likely that Week 52 will shift as well, next week.  Most of the
census divisions have reported the Week 52 data with a higher sample
size (than Week 51 presented with last week).  Hopefully, this will
lessen any shifts in the infectivity rates come next week.  We may
have to adjust the observational error variance estimate of the ILI+
data based on the sample size of the census division data (though this
will give no indication of spatial heterogeneities within these
relatively large geographies, if a particular locality, say Chicago,
hasn't reported at all).

This year, given the high levels of media coverage of influenza
activity, it is also possible that the GFT ILI estimates have become
biased high in the last few weeks.

\subsection{A Test}

We tested the effect of the shifted observations, by re-running the
Week 51 forecast--i.e. assimilation of observations through Week 51
and then forecast beginning December 23, 2012--however, instead of
using the census division infectivity rates posted December 28, 2012,
as performed previously \citep{Shaman-Karspeck-Lipsitch-2012:week51},
we used the observations posted January 4, 2013.  That is, we used the
latest observations, but dropped the Week 52 data, and ran the
forecasts after assimilating the Week 51 data.

When the results of these forecasts (not shown) are compared with
those generated last week (again the only difference is the altered
census division infectivity rates), the predicted peaks shift for 6 of
the 12 cities.  St. Louis, Seattle and Memphis are predicted to peak a
full week later; Los Angeles and Houston are predicted to peak a half
week later (e.g. from a Week 50 prediction to a Week 50-51); and New
York City is predicted to peak a half week earlier (from Weeks 51-52
to Week 51).

New York City is in the Mid-Atlantic census division for which the
infectivity rate doubled with the revised numbers.  Los Angeles,
however, is in the Pacific census division, which saw a very small
change.  In truth, the changes in the forecasts often are more subtle
and really reflect a change in the distribution of predictions within
each ensemble, such that the mode prediction changes.  We shall see
how important these effects are after the season when the numbers are
fully codified and we can run the forecasts again retrospectively and
determine if the forecast accuracy truly was degraded by these hiccups
in the data flow.

\section{2012-2013 Forecast}
\label{sec:actualfore}

The forecast methods are similar to those described in
\cite{Shaman-Karspeck-2012:forecasting}.  Based on the relationship
between prediction accuracy and ensemble spread of these retrospective
forecasts we can assign calibrated confidences to our current
predictions.  Two forecast types are presented: one run with an
absolute humidity (AH)-forced SIRS model; the other with no AH
forcing.  Additional documentation of earlier forecasts for this
season have also been posted
\citep{Shaman-Karspeck-Lipsitch-2012:week49,
  Shaman-Karspeck-Lipsitch-2012:week50,
  Shaman-Karspeck-Lipsitch-2012:week51}.

\subsection{Week 52 Forecast}
\label{subsec:actual52}

Table \ref{table:t2} presents the forecasts initiated after
assimilation of observations through Week 52.  The first forecast day
is December 30, 2012.  These forecasts use the AH-forced SIRS model
and were performed on January 4, 2013 with GFT ILI municipal estimates
and census division infectivity rates through Week 52 (the latter as
published online on January 4, 2013).

\begin{table}[t]
\caption{Summary of weekly model predictions at 12 select cities.  Weeks
   are labeled consecutively (Week 1 of 2013 is Week 53, etc.).
   Predictions were initiated at the end of Weeks 47-52.
   The range of prediction confidences, derived from municipal,
   regional and national calibrations, are given in parentheses.}\label{table:t2}
 \begin{center}
\resizebox{16cm}{!} {
 \begin{tabular}{ccccrrcrc}
   \hline\hline
   City & Week 52 &Week 51 & Week 50 & Week 49 & Week 48 & Week 47\\
   & Prediction &Prediction & Prediction & Prediction & Prediction & Prediction \\
   \hline
   Los Angeles &53 (35-95\%)&51-52 (35-60\%)&52 (50-95\%) & 51-52 (35-90\%) & 51-52 (20-55\%) & 51 (15-30\%) \\
   San Francisco &53 (35-60\%)&52 (25-45\%)& 52 (35-85\%) & 51-52 (25-40\%) & 51 (30-85\%) &
   50-51 (25-60\%)  \\
   Denver &52 (60-99\%)&52 (50-99\%)& 52 (20-60\%) & 52 (20-55\%) & 51-52 (0-55\%) & 51 (10-30\%) \\
   Miami &51 (65-99\%) &51-52 (30-99\%)& 51 (40-60\%) & 51 (40-99\%) & 50-51 (40-55\%) & 50-51 (0-45\%) \\
   Atlanta &49-50 (80-99\%)&49-50 (80-99\%)& 49 (80-99\%) & 49 (90-99\%) & 49 (80-95\%) & 49 (80-95\%) \\
   Chicago &50 (55-95\%)&50 (55-95\%)&49-50 (55-95\%) & 49 (55-95\%) & 49 (35-80\%) & 49 (35-80\%) \\
   St. Louis &51 (80-99\%) &50 (80-99\%)& 51 (85-99\%)  & 50-51 (80-99\%) & 50 (85-99\%) & 51 (30-90\%) \\
   New York City&52 (85-99\%) &51-52 (20-99\%) &52 (25-99\%) & 51 (25-99\%) & 52-53 (25-60\%) & 53-54
   (25-55\%)\\
   Memphis &51 (80-99\%)&50 (70-99\%)&51 (20-80\%) & 50 (20-80\%) & 50 (15-80\%) & 49-50 (15-55\%) \\
   Dallas &50 (40-70\%)&50 (65-95\%) &50 (65-90\%) & 49-50 (65-85\%) & 49 (50-75\%) & 49 (40-85\%) \\
   Houston & 50-51 (70-95\%) &50 (70-95\%)&50 (75-90\%) & 50 (50-60\%) & 50 (50-60\%) & 49 (50-85\%) \\
   Seattle &53 (50-90\%) &51-52 (20-60\%)&52-53 (0-55\%) & 52-53 (5-55\%) & 51-52 (5-55\%) & 51 (5-35\%) \\
   \hline
 \end{tabular}
}
 \end{center}
\end{table}

As a number of observations now indicate a later observed peak, the
city forecasts have shifted later.  Note that these ``forecasts'' are
in fact predictions that the peak is in the past, e.g. Week 51;
however, they are still a ``forecast'' in that there is a prediction
of no further increase of influenza incidence.

\begin{figure}[tbh]
\noindent\includegraphics[width=25pc,angle=0]{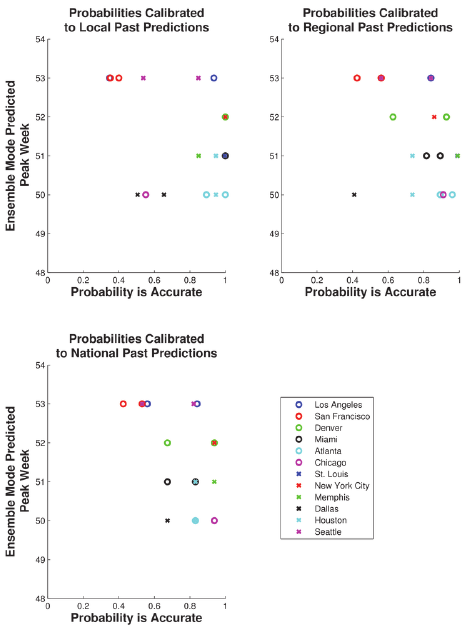}\\
\caption{Ensemble mode peak week predictions initiated December 30,
  2012, following assimilation of Week 52 observations, for 12 cities
  plotted as a function of probability/confidence calibrated from
  historical city, regional and national prediction accuracy.}
\label{fig:select_wk52fore_cal} 
\end{figure}

Atlanta is predicted to peak during Weeks 49-50, and Chicago and
Dallas are predicted to peak during Week 50 (Figure
\ref{fig:select_wk52fore_cal}).  At present, Week 51 is the observed
peak for Atlanta and Dallas (Figure \ref{fig:selcitytimeseries}),
which is within the $\pm 1$ accuracy of the forecast for the Dallas
forecasts, but Week 52 is greatest (thus far) for Chicago.
Consequently, the Chicago forecast is not accurate (in spite of the
high certainty, Table \ref{table:t2}).

\begin{figure}[tbh]
\noindent\includegraphics[width=18pc,angle=0]{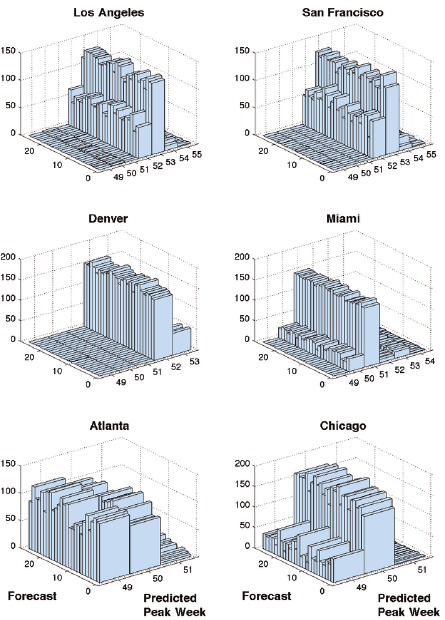}
\noindent\includegraphics[width=18pc,angle=0]{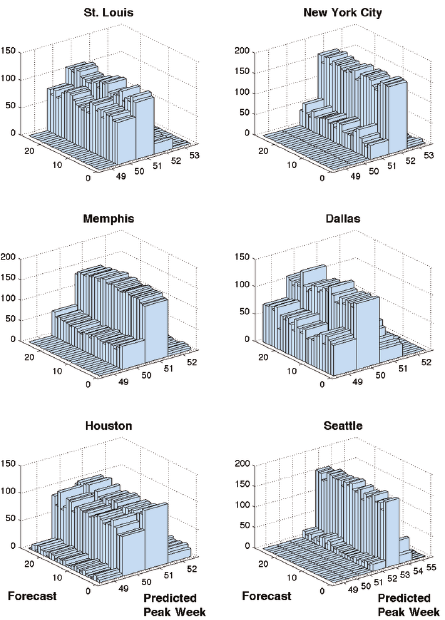}\\
\caption{Left) Histograms of the best ensemble start date trainings
  for forecasts made beginning the start of Week 53/1 (December 30,
  2012) for select cities.  The distributions show the ensemble spread
  among peak predictions.} 
\label{fig:select_wk52fore}
\end{figure}

Houston has shifted slightly with peak predictions of Week 50-51
(Figure \ref{fig:select_wk52fore}).  Presently, the peak is observed
for Week 51.

Miami, St. Louis and Memphis are now all predicted to have peaked
during Week 51.  This is a one-week shift later from the prior weeks
for St. Louis and Memphis, though both had early forecasts predicting
a Week 51 peak (Table \ref{table:t2}).  Miami appears to have peaked
during week 50, St. Louis during week 51 (by a smidge), and Memphis
also during week 51 (Figure \ref{fig:selcitytimeseries}).

Denver and New York City are predicted to have a Week 52 peak.  These
predictions are consistent with prior weeks.  Both Denver and New York
City showed continued increasing influenza activity, as measured by
our ILI+ metric, through Week 52 (Figure
\ref{fig:selcitytimeseries}).

Los Angeles, San Francisco, and Seattle have all shifted, and are now
predicting a peak during Week 53 (the week ending January 5, 2013).  

\subsection{Week 52 Forecast -- No AH}
\label{subsec:actual52noAH}

Forecasts initiated at the beginning of Week 53/1 (December 30, 2012,
after assimilation of Week 52 observations) using an SIRS model
without absolute humidity forcing show a few peak week prediction
shifts from the previous weeks (Table \ref{table:t3}).  Both Los
Angeles and San Francisco have predicted peaks during weeks 52-53 with
this model, and this is a half week later than the prediction made
last week (Week 51), and 1.5 weeks later than the Week 50 prediction.
In contrast, the predictions for Los Angeles and San Francisco with
the AH-forced model have been a bit more stable from week-to-week
(Table \ref{table:t2}).  This, of course, provides no assessment of
their accuracy, just that they have not shifted quite as much.

 \begin{table}[t]
   \caption{Summary of weekly model predictions at 12 select cities
     using an SIRS model without absolute humidity forcing.  Weeks
     are labeled consecutively (Week 1 of 2013 is Week 53, etc.).
     Predictions were initiated at the end of Weeks 48-52.
     The range of prediction confidences, derived from municipal,
     regional and national calibrations, are given in parentheses.}\label{table:t3}
 \begin{center}
\resizebox{16cm}{!} {
 \begin{tabular}{ccccrrcrc}
 \hline\hline
 City & Week 52 &Week 51 &Week 50 & Week 49 & Week 48 \\
    & Prediction &Prediction &Prediction & Prediction & Prediction \\
 \hline
  Los Angeles &52-53 (25-99\%) &52 (25-50\%)&51 (25-50\%) & 50-51 (25-50\%) & 50 (25-50\%) \\
  San Francisco &52-53 (20-85\%) &52 (30-50\%)& 51 (30-60\%) & 50-51 (30-60\%) & 50 (30-50\%) \\
  Denver &52 (90-99\%)&52 (50-99\%)& 52 (50-80\%) & 51-52 (40-85\%) & 51 (40-60\%)  \\
  Miami &50 (55-90\%)&50 (55-80\%& 50-51 (40-80\%) & 50 (10-99\%) & 50 (5-65\%) \\
  Atlanta &49 (40-99\%)&49-50 (80-99\%)&49 (80-99\%) & 49 (90-99\%) & 49 (25-95\%) \\
  Chicago &49-50 (55-99\%)&50 (45-95\%)&49 (55-95\%) & 49 (55-95\%) & 49 (25-65\%) \\
  St. Louis &51 (80-99\%)&50 (70-99\%)& 51 (80-95\%)  & 50 (80-95\%) & 50 (35-95\%) \\
  New York City &52 (80-99\%)&52 (30-60\%)&51-52 (30-60\%) & 51 (30-60\%) & 52-53 (25-60\%) \\
  Memphis &51 (90-99\%)&50 (70-99\%)&50 (45-99\%) & 50 (10-90\%) & 49-50 (15-55\%) \\
  Dallas &50-51 (15-85\%)&50 (80-95\%)&50 (65-90\%) & 49 (15-85\%) & 49 (40-80\%) \\
  Houston &50-51 (70-95\%)&50 (70-90\%)&50 (80-95\%) & 50 (60-70\%) & 49-50 (30-70\%) \\
  Seattle &53 (35-90\%)&52 (20-65\%)&51-52 (20-50\%) & 51-52 (20-45\%) & 51 (0-50\%)\\
 \hline
 \end{tabular}
}
 \end{center}
\end{table}

\begin{figure}[tbh]
\noindent\includegraphics[width=25pc,angle=0]{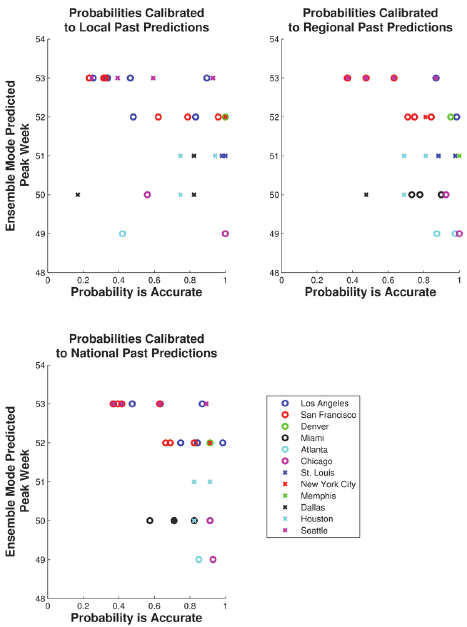}\\
\caption{Ensemble mode peak week predictions initiated December 30,
  2012, following assimilation of Week 52 observations using an SIRS
  model without AH forcing, for 12 cities plotted as a function of
  probability/confidence calibrated from historical city, regional and
  national prediction accuracy.}
\label{fig:select_wk52fore_noAH_cal}
\end{figure}

The other city forecasts with the SIRS model without AH forcing have
been more stable in recent weeks (perhaps with the exception of
Seattle (Figure \ref{fig:select_wk52fore_noAH_cal}).  These cities are
also fairly consistent with the AH-forced solutions (Table
\ref{table:t2}).

\bigskip
\textbf{Acknowledgments}

\bigskip

\myacknow

{}
{\clearpage}
\bibliographystyle{apalike} 
\bibliography{week48bib}


\end{document}